\documentstyle[12pt]{article}
\begin{document}
hep-th/9306155\\
Mathematical Preprint Series, Report 93-11, University of Amsterdam
\vskip 15mm
\title{Linear $r$-matrix algebra for classical \\
separable systems}

\vskip 15mm
\author{ J.C.Eilbeck${}\sp a$, V.Z.Enol'skii${}\sp b$,
V.B.Kuznetsov${}\sp {c{}1}${} and  A.V.Tsiganov${}\sp d$}
\footnotetext[1]{This author was supported by the
National Dutch Science
Organization (NWO) under the Project \# 611--306--540.}

\maketitle

{\it ${}\sp a$Department of Mathematics, Heriot-Watt University,
\\ Riccarton, Edinburgh EH14 4AS, Scotland \\
e-mail: chris@cara.ma.hw.ac.uk

${}\sp b$Department of Theoretical Physics, Institute of Metal Physics,
\\ Vernadsky str. 36, Kiev-680, 252142, Ukraine \\
e-mail: vbariakhtar@gluk.apc.org

${}\sp c$Department of Mathematics and Computer Science, University of
Amsterdam, \\Plantage Muidergraht 24, 1018 TV Amsterdam, The Netherlands
\\ e-mail: vadim@fwi.uva.nl

${}\sp d$Department of Earth Physics, Institute for Physics, University
of St Petersburg,\\ St Petersburg 198904, Russia \\
e-mail: kuznetso@onti.phys.lgu.spb.su}

\newcommand{\PB}{\stackrel{\textstyle\otimes}{,}}
\renewcommand{\theequation}{\arabic{section}.\arabic{equation}}
\newtheorem{theorem}{Theorem}
\newtheorem{prop}{Proposition}
\newtheorem{cor}{Corollary}
\vskip 10mm
\begin{abstract}
We consider a hierarchy of the natural type Hamiltonian systems of $n$
 degrees of freedom
 with polynomial potentials
separable in general ellipsoidal and general paraboloidal coordinates.
 We give a
Lax representation in terms of $2\times 2$ matrices for the whole hierarchy
 and construct
the associated linear $r$-matrix algebra with the $r$-matrix dependent on
the dynamical
variables. A Yang-Baxter equation of dynamical type  is proposed. Using
the method of variable separation we provide the integration of the
systems in classical mechanics conctructing the separation equations
and, hence, the explicit form of action variables. The quantisation problem
is discussed  with
the help of the separation variables. \end{abstract}

\vskip 10mm
{\bf Key words:} separable systems, linear $r$-matrix algebras, classical
 Yang-Baxter equation, Lax representation, Liouville integrable systems

\vskip 5mm
{\bf AMS classification:} 70H20, 70H33, 58F07

\vskip 10mm
\pagebreak
\section{Introduction}
\setcounter{equation}{0}

The method of separation of variables in the Hamilton-Jacobi equation,
\begin{equation}H(p_1,\ldots,p_n,x_1,\ldots,x_n)=E,\quad p_i=
{\partial W\over \partial x_i},
\quad i=1,\ldots n,\label{hje}\end{equation}
is one of the most powerful methods for the construction of action for the
 Liouville
integrable systems of classical mechanics \cite{ar74}. We consider below
the systems of the {\it natural\,} form described by the Hamiltonian
\begin{equation}H={1\over2}\sum_{i=1}^np_i^2+V(x_1,\ldots,x_n),\quad p_i,x_i
 \in \bf R \, .
\label{ham}\end{equation}
The separation of variables means the solution of the partial differential
 equation
(\ref{hje}) for the action function $W$ in the following additive form
\[W=\sum_{i=1}^nW_i({\mu}_i;H_1,\ldots,H_n),\quad H_n=H,\]
where ${\mu}_i$  will be
called {\it separation
variables\,}. Notice that the partial functions $W_i$ depend only on
their separation
variables ${\mu}_i$, which define new set of variables instead of the
 old ones \{$x_k$\}, and on the set of constants of motion, or
{\it integrals of motion\,}, $\{H_k\}$. In the sequel we shall speak
 about {\it coordinate\,} separation where the
separation variables \{${\mu}_i$\}  are the functions of the coordinates
 \{$x_k$\} only. (The general
change of variables may include also corresponding momenta \{$p_k$\}).

For a free particle $(V=0)$, the complete classification of
all orthogonal
coordinate systems in which the Hamilton-Jacobi equation (\ref{hje})
admits the separation
of variables is known, they are generalized $n$-dimensional ellipsoidal and
paraboloidal coordinates
\cite{ka86,mi86} (see also the references therein). It is also
known that the Hamiltonian
systems (\ref{ham}) admitting an orthogonal coordinate separation with
$V\neq 0$ are separated only in the {\it same \,} coordinate systems.

The modern approach to finite-dimensional integrable systems uses the
 language of the
representations of $r$-matrix algebras \cite{ks82,rs87,sk84,sk89}.
The classical method
of separation
of variables can be formulated within this language dealing with the
representations of
linear and quadratic $r$-matrix algebras \cite{kuz92,ku92,sk84,sk92}.
For the $2\times 2$ $L$-operators the recipe consists of considerating
 the zeros of one of the
off-diagonal elements as the separation variables (see also generalization
 of this
approach to the higher dimensions of $L$-matrix \cite{sk92}). For $V=0$
in \cite{kuz92,ku92}
$2\times 2$ $L$-operators were given satisfying the standard linear
 $r$-matrix algebra \cite{ks82,rs87},
\begin{equation}\{L_1(u),L_2(v)\}=[r(u-v),L_1(u)+L_2(v)],
\quad r(u)={1\over u}\left(\begin{array}
{llll}1&0&0&0\\0&0&1&0\\0&1&0&0\\0&0&0&1\end{array}\right),
\label{stal}\end{equation}
and the link with the separation of variables method was elucidated.
Here in (\ref{stal}) we use
the familiar notations for the tensor products of $L(u)$ and
 $2\times 2$ unit matrix
$I$, \,$L_1(u)=L(u)\otimes I$, \,$L_2(v)=I\otimes L(v)$.

In the present article we construct $2\times 2$
$L$-operators for the systems (\ref{ham}) being separated in
the generalized ellipsoidal and paraboloidal coordinates.
In the case when the degree $N$ of the potential $V$ is equal
to $1$ or $2$ the associated linear $r$-matrix algebra appears to
be the standard one (\ref{stal}). In the case $N>2$ the algebra
is of the form
\begin{equation}\{L_1(u),L_2(v)\}=
[r(u-v),L_1(u)+L_2(v)]+[s(u,v),L_1(u)-L_2(v)],\end{equation}
with $s(u,v)=\alpha_N(u,v)\,\sigma_-\otimes\sigma_-$, where
$\sigma_-=\sigma_1-i\,\sigma_2$ and $\sigma_i$ are the Pauli matrices,
and $\alpha_N(u,v)$ is the function which equals $1$ for $N=3$ and
depends on the dynamical variables $\{x_k,p_k\}_{k=1}^n$ for $N>3$.

The study of completely integrable systems admitting a
classical $r$-matrix Poisson structure with the $r$-matrix
dependent on dynamical variables has attracted some attention
\cite{bv90,fm91,ma85}. It is remarkable that the celebrated
Calogero-Moser system, whose complete integrability was shown
a number of years ago (c.f. \cite{op81}), has been found only recently
to possess a classical $r$-matrix of dynamical type \cite{at92}.

Below we recap briefly how to get the $2\times 2$ $L$-operators for the
separable systems (\ref{ham}) without potential $V$ \cite{kuz92,ku92}.
Let us consider the direct sum of the Lie algebras each of rank 1:
${\cal A}={}\oplus^n_{k=1}{\rm so}_{k}(2,1)$.
Generators $\vec s_{k}\in {\bf R}^3,\, k=1,\ldots,n$ of the $\cal A$
algebra satisfy the following Poisson brackets:
\begin{equation}\{s^i_{k},s^j_{m}\}=\delta_{k m} \,
\varepsilon_{ijl} \, g_{ll} \, s^l_{k},\quad g={\rm diag}(1,-1,-1).
\label{pb}\end{equation}
Everywhere below, we will imply the $g$ metric to calculate the norm and
scalar product of the vectors ${\vec s}_{i}$:
\begin{eqnarray}{\vec s}_{i}^{\,2}&&\equiv({\vec s}_{i},{\vec s}_{i})=
({ s}_{i}^1)^2-({ s}_{i}^2)^2-({ s}_{i}^3)^2,\nonumber\\
&&({\vec s}_{i},{\vec s}_{k})=
{ s}_{i}^1{ s}_{k}^1-{ s}_{i}^2{ s}_{k}^2-
{ s}_{i}^3{ s}_{k}^3.\nonumber\end{eqnarray}
Let us fix the values of the Casimir elements of the ${\cal A}$ algebra:
${\vec s}_{i}^{\,2}=c_{i}^2$, then variables ${\vec
s}_{i}$ lie on the direct product of $n$ hyperboloids in ${\bf R}^3$.
Let $c_{i}\in\bf R$ and choose the upper sheets of these
double-sheeted hyperboloids. Denote the obtained manifold as $K_n^+$.
We will denote by {\it hyperbolic Gaudin magnet\,} \cite{ga83} integrable
Hamiltonian system on $K_n^+$ given by $n$ integrals of
motion $H_{i}$ which are in involution with respect to the bracket
(\ref{pb}),
\begin{equation}H_{i}=2{\sum_{k=1}^n}'
{({\vec s}_{i},{\vec s}_{k})\over e_{i}-e_{k}},\quad
\{H_{i},H_{k}\}=0, \quad e_{i}\neq e_{k}\quad {\rm if} \quad i\neq k.
\label{ga}\end{equation}
To be more exact one has to call this model an $n$-site so(2,1)-$XXX$
Gaudin magnet.
Notice that all the $H_{i}$ are quadratic
functions on generators of the $\cal A$ algebra and the following equalities
 are valid
\[\sum_{i=1}^nH_{i}=0,\quad\sum_{i=1}^ne_{i}H_{i}=
 \vec J^{\,2}-\sum_{i=1}^n c_{i}^2,\]
where the new variable $\vec J=\sum_{i=1}^n \vec s_{i}$\, is introduced
which is the total sum of the hyperbolic momenta $\vec s_{i}$.
The components of the vector $\vec J$ obey so(2,1) Lie algebra with
respect to the bracket (\ref{pb}) and are in involution with all
the $H_{i}$. The complete set of involutive integrals of motion is
 provided by the
following choice: $H_{i}, \vec J^{\,2}$ and, for
example, $(J^3)^2$. The integrals (\ref{ga}) are generated by the
$2\times 2$ $L$-operator
(as well as the additional integrals $\vec J$\,)
\begin{eqnarray}L(u)&=& \sum_{j=1}^n
{1\over u-e_{j}} \left(\begin{array}{cc}is^3_{j}&-(s^1_{j}-s^2_{j})\\
-(s^1_{j}+s^2_{j})&-is^3_{j}\end{array}\right),\nonumber\\
&&{\rm det} \,L(u)=-\sum_{j=1}^n{\left({H_{j}\over u-e_{j}}
+{c^2_{j}\over(u-e_{j})^2}\right)},\label{lax1}\end{eqnarray}
satisfying the standard linear $r$-matrix algebra (\ref{stal}).
Let $c_{i}=0,\,i=1,\ldots,n$, then the hyperboloids $\vec s_i^{\,2}=c_{i}^2$
 turn into cones.
The manifold $K_n^+$ admits
in this case the following parameterization $(p_{i},x_{i}\in {\bf R})$:
\begin{equation}s^1_{i}={p_{i}^2+x_{i}^2\over 4},
\quad s^2_{i}={p_{i}^2-x_{i}^2\over 4},\quad
s^3_{i}={p_{i}\,x_{i}\over 2},\label{ss}\end{equation}
where the variables $p_{i}$ and $x_{i}$ are canonically conjugated.
Using the isomorphism (\ref{ss}) the complete classification of the
separable orthogonal
coordinate systems was provided in \cite{kuz92,ku92} by means of the
corresponding  $L$-operators satisfying the standard linear $r$-matrix
algebra (\ref{stal}). The starting point for our investigation are these
$L$-operators written for the cases of
free motion on a sphere and in the  Euclidean space.

The paper is organized as follows. In Section 2 we describe the
classical Poisson structure associated with the hierarchy of natural
 type Hamiltonians separable in the three coordinate systems---spherical
 (for motion on a sphere), and general ellipsoidal and paraboloidal
 (for $n$-dimensional Euclidean motion) coordinates. This structure is given
in terms of the linear $r$-matrix formalism, providing new example
 of the dynamical dependence of the
$r$-matrices. We also introduce analogue of Yang-Baxter equation
for our dynamical $r$-matrices. In the Section 3 we derive the Lax
representation for all the hierarchy as a
consequence of the $r$-matrix representation given in the Section 2.
Section 3 deals also with the variable separation material. The question
of quantisation of the considered systems is briefly
discussed.

\section{Classical Poisson structure}
\setcounter{equation}{0}
Let us consider the following ansatz for the $2\times2$ $L$-operator
\begin{equation}L_{N}(u)=\left(\matrix{ A(u) & B(u)\cr
   C_{N}(u)&-A(u)}\right)\label{l} \end{equation}
where
\begin{eqnarray}
B(u)&=&\varepsilon -\sum_{i=1}^n{x_i^2\over u - e_i}\,, \quad
\varepsilon = 0,\,1,\, {\rm or} \quad 4(u-x_{n+1}+B), \label{uz}\\
A(u)&=&\dot\varepsilon+{\frac12}\sum_{i=1}^n{x_i\, p_i\over u - e_i}\,,
\label{Vz}\\
C_{N}(u)&=& \sum_{i=1}^n{p_i^2\over u - e_i}-V_N(u),\quad
V_N(u) = \sum_{k=0}^N{\cal V}_k\,u^{N-k}.\label{vnz} \end{eqnarray}
Here the $x_i,p_j$ are canonically conjugated variables
($\{p_i,x_j\}=\delta_{ij}$), \,${\cal V}_k$ are indeterminate
functions of the $x$-variables; $B$ and $e_i$ are
non-coincident real constants. Note that dot over $\varepsilon$ means
differentiation
by time, and for natural Hamiltonian (\ref{ham}) one has
$\dot x_{n+1}= p_{n+1}$.
\begin{theorem} Let the curve $\det(L(u)-\lambda I)=0$ for the $L$-operator
 (\ref{l}) have the
form
\begin{eqnarray}
\lambda^2-A(u)^2-B(u) C_N(u)&=&\lambda^2+
\varepsilon\, u^N-\sum_{i=1}^n{H_i\over u -
e_i}= 0,\nonumber\\&&{\rm for}\quad
	\varepsilon=0,1,\quad {\rm and }\label{curve1} \\
\lambda^2-A(u)^2-B(u) C_N(u)&=&\lambda^2+16u^{N-2}(u+B)^2+8H
-\sum_{i=1}^n{H_i\over u - e_i}=
 0,\nonumber\\&&{\rm for}\quad \varepsilon=4(u-x_{n+1}+B)\,,\label{curve2}
\end{eqnarray}
with some integrals of motion $H_i$ and $H_i,\,H$ in the case of
(\ref{curve2}).
Then the following recurrence relations for ${\cal V}_k$ are valid
\begin{eqnarray}
{\cal V}_k&=&\sum_{i=1}^n x_i^2\sum_{j=0}^{k-1}{\cal V}_{k-1-j}e_i^j,
	\quad {\cal V}_0=1,\nonumber\\&&
	{\rm for}\quad \varepsilon=0,1;\label{ucal1} \\
{\cal V}_k&=&(x_{n+1}+B){\cal V}_{k-1}+{1 \over2}
\sum_{i=1}^n x_i^2\sum_{j=1}^{k-1}{\cal V}_{k-1-j}e_i^j,
	\quad {\cal V}_0=0, \nonumber\\&&
	{\rm for}\quad \varepsilon=4(u-x_{n+1}+B).\label{ucal2}
\end{eqnarray}
The explicit formulae for the integrals $H_i$ have the form
\begin{eqnarray}
H_i&=&-{\sum_{j=1}^n}'\frac{M_{ij}^2}{e_i-e_j} + \varepsilon\cdot p^2_i +
	x_i^2\sum_{k=0}^N{\cal V}_k e_i^{N-k},\quad {\rm for}\quad
 \varepsilon=0,1, \label{h1}\\
H_i&=&2x_i^2\sum_{j=1}^{N-1}(-1)^{j-1}e_i^j\,{\cal V}_{N-j}
+4p_{n+1}\,p_i\,x_i-p_i^2(e_i+4x_{n+1}-4B)\nonumber\\&
+&\sum_{j\neq i}{M_{ij}^2\over e_i-e_j},\quad
	\quad{\rm for}\quad \varepsilon=4(u-x_{n+1}+B),\label{h2}
\end{eqnarray}
where $M_{ij}=x_ip_j-x_jp_i$. The Hamiltonians $H$ are given by
\begin{eqnarray}
H&\equiv&\sum_{i=1}^nH_i=\varepsilon\cdot\sum_{i=1}^n p_i^2+{\cal V}_{N+1},
	\quad{\rm for}\quad \varepsilon=0,1, \label{HH1}\\
H&=&{1\over
2}\sum_{i=1}^{n+1} p_i^2 +{\cal V}_N,
	\quad{\rm for}\quad \varepsilon=4(u-x_{n+1}+B).\label{HH2}
\end{eqnarray}
\end{theorem}
{\bf Proof} is straightforward and based on direct computations.

We remark, that the above recurrence formulae for the potentials can
be written in differential form. In particular, for the paraboloidal
coordinates we have
\begin{eqnarray}
{\partial {\cal V}_N\over \partial
x_i}&=&{1\over2}{\partial {\cal V}_{N-1}\over \partial x_{n+1}} x_i
-A_i {\partial {\cal V}_{N-1}\over \partial
x_i},\quad i=1,\ldots,n,\label{equ1}\\ {\partial {\cal V}_N\over \partial
x_{n+1}}&=&{\cal V}_{N-1} +{1\over2}\sum_{i=1}^n x_i {\partial {\cal
V}_{N-1}\over \partial x_i}+(x_{n+1}-B){\partial {\cal V}_{N-1}\over
\partial x_{n+1}}\,.\label{equ2}
\end{eqnarray}

Notice that the case of $\varepsilon=0$ is connected with the ellipsoidal
 coordinates on
a sphere and two other cases $\varepsilon =1, \,4(u-x_{n+1}+B) $ describe
 the ellipsoidal and
paraboloidal coordinates in the Euclidean space, respectively (see Section
 3.2 and \cite{kuz92,ku92}
for more details). Recall that we study now the motion of a particle on
 these manifolds under
the external field with the potential $V$ that could be any linear
 combination of the
homogeneous ones ${\cal V}_k$.

Now we are ready to describe the linear algebra for the $L$-operator
 (\ref{l}).
\begin{theorem}
Let the $L$-matrix be of the form (\ref{l}) and satisfy the conditions
of the Theorem 1,
then the following algebra is valid for its entries
\begin{eqnarray}
\{B(u),B(v)\}&=&\{A(u),A(v)\}=0, \label{uu}\\
\{C_N(u),C_N(v)\}&=&-4\,\alpha_N(u,v)\,(A(u)-A(v)), \label{ww}\\
\{B(u),A(v)\}&=&\frac2{u-v}(B(u)-B(v)), \label{uv}\\
\{C_N(u),A(v)\}&=&-\frac2{u-v}(C_N(u)-C_N(v))-2\,\alpha_N(u,v)\,B(v),
 \label{wv}\\
\{B(u),C_N(v)\}&=&-\frac4{u-v}(A(u)-A(v)), \label{uw} \end{eqnarray}
where the function $\alpha_N(u,v)$ has the form
\begin{eqnarray}
\alpha_N(u,v)&=&\frac{Q_N(u)-Q_N(v)}{u-v}=
\sum_{k=1}^{N} Q_k \frac{u^k-v^k}{u-v},\label{alpha}\\
Q_N(u)&=&\sum_{k=0}^N Q_k\, u^k,\quad
Q_k = \sum_{m=0}^{k} {\cal V}_{m} \,{\cal V}_{k-m}.
\nonumber\end{eqnarray}
\end{theorem}
{\bf Proof} is based on the recurrence relations (\ref{ucal1}),
(\ref{ucal2}).

We remark that for the paraboloidal coordinates the following formula
is valid
\begin{equation}Q(u)=u^{N-2}-{1\over
  2}\sum_{k=0}^{N-3}{\partial {\cal
  V}_{N-k-1}\over \partial x_{n+1}}\,,\label{q}\end{equation}
therefore in this case we have
\begin{equation}Q(u)={1\over 4}{\partial
C(u)\over \partial x_{n+1}}\,.\label{qc}\end{equation}

The algebra (\ref{uu})-(\ref{uw}) can be rewritten in the matrix
form as linear
$r$-matrix algebra
\begin{equation}
\{L_1(u),L_2(v)\}=[r(u-v),L_1(u)+L_2(v)]+[s_N(u,v),L_1(u)-L_2(v)],
\label{poi}
\end{equation}
 using $4\times 4$ notations
$L_1(u)= L(u)\otimes I,\,L_2(v)=I\otimes L(v)$; the matrices
$r(u-v)$ and $s_N(u,v)$ are given by
\begin{eqnarray}r(u-v)=\frac2{u-v}\,P,
\quad P=\left(\begin{array}{llll}
						1&0&0&0\\
						0&0&1&0\\
						0&1&0&0\\
						0&0&0&1
					\end{array}\right),
\label{rs}\\
s_N(u,v)=2\,\alpha_N(u,v)\,\sigma_-\otimes\sigma_-,\quad
\sigma_-=\left(\begin{array}{ll}0&0\\1&0\end{array}\right).
\nonumber\end{eqnarray}
The algebra (\ref{uu})-(\ref{uw}) or (\ref{poi})-(\ref{rs})
contains all the information
about the system under consideration. From it there follows
the involutivity
of the integrals of motion. Indeed, the determinant
$d(u)\equiv \det L(u)$ is the generating
function for the integrals of motion and it is simply to show that
\begin{equation}\{d(u),d(v)\}=0.\label{int}\end{equation}
In particular, the integrals $H_i$ (\ref{h1}), (\ref{h2}) are the
residues of the function $d(u)$:
$$H_i={\rm res}|_{u=e_i}\,d(u),\quad i=1,\ldots,n.$$
The Hamiltonians $H$ (\ref{HH1}),(\ref{HH2}) appear to be a residue
at infinity.
Let us rewrite the relation (\ref{poi}) in the form
\begin{equation}
\{L_1(u), L_2(v)\} = [d_{12}(u,v),L_1(u)]-[d_{21}(u,v)L_2(v)],
\label{rsal2}\end{equation}
with $d_{ij}=r_{ij}+s_{ij},\,d_{ji}=s_{ij}-r_{ij}$ at $i<j$.
\begin{theorem} The following equations (dynamical Yang-Baxter equations)
are valid
for the algebra (\ref{rsal2})
\begin{eqnarray}
&&[d_{12}(u,v),d_{13}(u,w)]+[d_{12}(u,v),d_{23}(v,w)]+
[d_{32}(w,v),d_{13}(u,w)]+\nonumber\\
&&+\{L_2 (v), d_{13}(u,w)\}-\{L_3 (w), d_{12}(u,v)\} +\nonumber\\
&&\quad+[c(u,v,w), L_2 (v)-L_3 (w)]=0,\label{ybe4} \end{eqnarray}
where $c (u,v,w)$ is some matrix dependent on dynamical
variables. Other two equations are obtained from (\ref{ybe4}) by
cyclic permutations.\end{theorem}
{\bf Proof} Let us write the Jacobi identity as
\begin{eqnarray} &&\{L_1 (u), \{L_2 (v), L_3 (w)\}\}+
\{L_3 (w),\{L_1 (u), L_2 (v)\}\}\nonumber\\
&&\quad+\{L_2 (v),\{L_3 (w), L_1 (u)\}\}=0\label{Jaid} \end{eqnarray}
with $L_1 (u)=L (u)\otimes I \otimes I$, $L_2 (v)= I \otimes L (v)\otimes I$,
$L_3 (w)= I \otimes I\otimes L (w)$. The extended form of the (\ref{Jaid})
reads \cite{ma85},
\begin{eqnarray}
&&[L_1 (u),[d_{12}(u,v),d_{13}(u,w)]+[d_{12}(u,v),d_{23}(v,w)]+\nonumber\\
&&+[d_{32}(w,v),d_{13}(u,w)]]+\label{ybe3}\\
&&+[L_1 (u), \{L_2 (v), d_{13}(u,w)\}-\{L_3 (w), d_{12}(u,v)\}] +\nonumber\\
&&+\quad\mbox{ cyclic permutations}=0.\nonumber \end{eqnarray}
Further we restrict ourselves proving the (\ref{ybe4}) only
in the paraboloidal case (other cases
can be handled in a similar way).
To complete the derivation of (\ref{ybe4}) we shall prove the
following equality for all the members of the hierarchy
\begin{eqnarray} \{L_2 (v), s_{13}(u,w)\}& -&\{L_3 (w), s_{12}(u,v)\}
= 2\beta_N(u,v,w)[P_{23},s_{13}+s_{12}]\nonumber\\&-&{\partial
\beta_N(u,v,w)\over \partial x_{n+1}}[s,L_2 (v)-L_3
(w)]\label{ybe1}\end{eqnarray} (with cyclic permutations). In
(\ref{ybe1}) the matrix $s=\sigma_-\otimes \sigma_-\otimes \sigma_-$
and
\begin{equation}\beta_N(u,v,w)={Q_N(u)(v-w)+Q_N(v)(w-u)+Q_N(w)(u-v)
\over (u-v)(v-w)(w-u)}.\label{hh}\end{equation}
In the extended form (\ref{ybe1}) can be rewritten as
\begin{eqnarray}\{Q(u),Q(v)\}&=&\{B(u),Q(v)\}=0,\label{qq}\\
\{A(u),Q(v)\}&=&4\,\alpha_N(u,v)-{1\over 2}{\partial
\alpha_N(u,v)\over \partial x_{n+1}}B(u),\label{vq}\\
\{A(w),\alpha_N(u,v)\}&=&{4\over u-v}(\alpha_N(w,u)-\alpha_N(w,v))
\nonumber\\
&-&{1\over 2} {B(w)\over u-v}\left({\partial \alpha_N(w,v)\over \partial
x_{n+1}}-{\partial \alpha_N(w,u)\over \partial
x_{n+1}}\right),\label{wp}\\
\{Q(u),C(v)\}&+&\{C(u),Q(v)\}={\partial \alpha_N(u,v)\over \partial
x_{n+1}}(A(u)-A(v)).\label{wq}\end{eqnarray}
The equality (\ref{qq}) is trivial and equation (\ref{vq}) is derived
by differentiating (\ref{wv}). Equation
(\ref{wp}) follows from the definition of $Q(u)$ and (\ref{vq}).
To prove (\ref{wq}) we write it using the explicit form of $C_N(u)$
and $A(u)$ as follows \begin{eqnarray}
&&\sum_{i=1}^n{p_i\over (v-e_i)(u-e_i)}\left((u-e_i){\partial
\alpha_N(w,u)\over \partial x_i}-(v-e_i){\partial \alpha_N(w,v) \over
\partial x_i}\right)\nonumber\\ &&={1\over 2}\sum_{i=1}^n{p_ix_i\over
(u-e_i)(v-e_i)} \left({\partial \alpha_N(w,u)\over \partial
x_{n+1}}-{\partial \alpha_N(w,v)\over \partial
 x_{n+1}}\right).\label{10}\end{eqnarray} Using  the identity
 \[u{w^k-u^k\over w-u}-v{w^k-v^k\over
w-v}={w^{k+1}-u^{k+1}\over w-u}-{w^{k+1}-v^{k+1}\over w-v},\]
and the recurrence relation (\ref{equ1}), we find that the equality
(\ref{10}) is valid. Therefore the equations (\ref{ybe4}) follow
with the matrix $c(u,v,w)=\partial
\beta(u,v,w)/\partial x_{n+1}\,\sigma_-\otimes\sigma_-\otimes\sigma_-$.
The proof is completed.

We remark that validity of the equations
(\ref{ybe4}) with an arbitrary matrix $c (u,v,w)$ is sufficient for the
validity of (\ref{Jaid}) and, therefore, (\ref{ybe4}) can be interpreted
as some dynamical classical Yang-Baxter equation, i.e. the associativity
condition for the linear $r$-matrix algebra.
 These equations have an
extra term $[c,L_i-L_j]$ in comparison with the extended Yang-Baxter
equations in \cite{ma85}.

We would like to emphasize that all statements of the Section can be
generalized to the following form of the potential term $V_N(u)$
in (\ref{vnz})
$$V_{MN}(u)=\sum_{k=-M}^N f_k\,{\cal V}_k \,u^{N-k},\quad f_k\in {\bf C}$$
that corresponds to the linear combinations of homogeneous terms
${\cal V}_k$ as potential $V$
and also includes the negative degrees to separable potential.
See the end of Section 3.1 for more details.

\section{Consequences of the $r$-matrix representation}
\setcounter{equation}{0}
\subsection{Lax representation}
Following the article \cite{bv90} we can consider the Poisson structure
(\ref{rsal2}) for the powers of the $L$-operator
\begin{equation}\{(L_1 (u))^k,(L_2 (v))^l\} =
[d^{(k,l)}_{12}(u,v),L_1 (u)]-[d^{(k,l)}_{21}(u,v)L_2 (v)],
\label{rsal3}\end{equation}
with \begin{eqnarray}
&&d^{(k,l)}_{ij}(u,v)=\label{dd}\\&&
\quad\sum_{p=0}^{k-1}\sum_{q=0}^{l-1}(L_1 (u))^{k-p-1}
(L_2 (v))^{k-q-1}d _{ij}(u,v)(L_1 (u))^{p}(L_2 (v))^{q}.
\nonumber\end{eqnarray}
As an immediate consequence of (\ref{rsal3})-(\ref{dd}) we
obtain that the
conserved quantities $H$, $H_i$ are in involution. Indeed, we have
\begin{equation}
\{{\rm Tr}(L_1 (u))^2,{\rm Tr}(L_2 (v))^2\}={\rm Tr}\{(L_1 (u))^2 ,
 (L_2 (v))^2\},
\end{equation}
and after applying the equality (\ref{rsal3}) at
$k=l=2$ to this equation and taking the trace, we
obtain the desired involutivity.
Further, let us define differentiation by time as follows:
\begin{equation}\dot L (u)={d\over dt}L(u)=
{\rm Tr}_2\{L_1 (u),(L_2 (u))^2\},
\label{lax}\end{equation}
where the trace is taken over the second space. Applying the equation
(\ref{rsal3}) at $k=1,l=2$ to (\ref{lax}), we obtain the Lax
representation
in the form $\dot L (u)=[M (u), L (u)]$ with the matrix $M (u)$ given as
\begin{equation}M (u)=2\lim_{v\rightarrow u}
{\rm Tr}_2 L_1 (v)(r(u-v)-s (u,v)). \end{equation}
After the calculation in which we take into account the asymptotic
 behaviour of the $L$-operator
(\ref{l}), we obtain the following explicit Lax representation:
\begin{eqnarray}\dot L(u)&=&[M(u),L(u)],\nonumber\\
L(u)&=&\left(\matrix{ A(u)&B(u)\cr C_N(u)&-A(u)}\right),
\quad M(u)=\left(\matrix{ 0&1\cr
Q_N(u)&0}\right),\label{Lax}\end{eqnarray}
where $Q_N(u)$ was defined by the equations (\ref{alpha}). Lax
representations
for the higher flows can be obtained in a similar way.

It follows from the equation (\ref{Lax}) that
\[A(u)=-\frac12 \dot{B}(u),
\quad C_N(u)=-\frac12 \ddot{B}(u)-B(u)Q_N(u),\]
so our $L$-matrix can be given in the form
\begin{equation}
L(u)=\left(\begin{array}{ll}-\frac12 \dot{B}(u)&B(u) \\
-\frac12 \ddot{B}(u)-B(u)Q_N(u)&\frac12 \dot{B}(u)\end{array}\right).
\label{plax}\end{equation}
The equations of motion, which follow from (\ref{Lax}) with
the $L$-matrix from
(\ref{plax}), have the form
\begin{equation}{\cal B}_1[Q_N]\cdot B(u)=0, \label{kdv}\end{equation}
where
\[{\cal B}_1[Q_N]\equiv \frac14 \partial^3+\frac12\{\partial,Q_N\},
\qquad\partial\equiv\frac{d}{dt},\]
with curly brackets standing for the anticommutator.
Operator ${\cal B}_1$ is the Hamiltonian operator of the first
Hamiltonian
structure for the coupled KdV equation \cite{ar92,arw92}.
Equation (\ref{kdv}), considered as one for the unknown function $B(u)$,
was solved in the three cases (\ref{uz}),
\[B(u)= \varepsilon -\sum_{i=1}^n{x_i^2\over u - e_i}\,,
\qquad\varepsilon=0,1,\,4(u-x_{n+1}+B),\]
in \cite{ar92} and \cite{arw92}. General solution of this
equation as one for the
$Q(u)$ has the form
\begin{equation}Q_{MN}(u)=\sum_{k=-M}^N f_k \,Q_k\, u^k,
\qquad f_k\in {\bf C}, \label{Qmn}\end{equation}
where the coefficients $Q_k$ are defined from the generating
function $\tilde{Q}(u)$
\begin{equation}
\tilde{Q}(u)\equiv B^{-2}(u)=\sum_{k=-\infty}^{+\infty} Q_k u^k.
\label{Qz}\end{equation}
Recall that we can write the element $C_N(u)$ of the $L$-matrix
(\ref{l}) in two different forms
(using $Q$ or $V$ functions)
\[C_N(u)=-\frac12 \ddot{B}(u)-B(u)Q_N(u)=-\sum_{i=1}^n{p_i^2
\over u - e_i}-V(u)\]
where function $V(u)=\sum_{k=0}^N{\cal V}_k\,u^{N-k}$ was
defined in (\ref{vnz}). The general form of the function $V(u)$
is as follows:
\begin{equation}V_{MN}(u)=\sum_{k=-M}^N f_k \,{\cal V}_k \,u^k,
\qquad f_k\in {\bf C}, \label{vmn}\end{equation}
where coefficients ${\cal V}_k$ are defined by the generating
function $\tilde{V}(u)$
\begin{equation}\tilde{V}(u)\equiv B^{-1}(u)=
\sum_{k=-\infty}^{+\infty} {\cal V}_k\, u^k.\label{vz}\end{equation}
Potentials ${\cal V}_k$ are connected with coefficients $Q_k$. Indeed,
using generating functions
(\ref{Qz}) and (\ref{vz}), we have
\[\tilde{Q}=B^{-2}(u)=\tilde{V}(u)\cdot\tilde{V}(u)=
\left({\sum_{k=-\infty}^{+\infty} {\cal V}_k\, u^k}\right)
\left({\sum_{k=-\infty}^{+\infty} {\cal V}_k \,u^k}\right)=
\sum_{k=-\infty}^{+\infty} u^k \sum_{j=0}^{k} {\cal V}_k\,
{\cal V}_{k-j}, \]
and, therefore, $Q_k=\sum_{j=0}^{k} {\cal V}_k \,{\cal V}_{k-j}$.
Thus we have recovered the formula (\ref{alpha}) for the $s$-matrix.

\subsection{Separation of Variables}
Let $K$ denote the number of degrees of freedom: $K=n-1$ for ellipsoidal
coordinates on a sphere, $K=n$ for ellipsoidal coordinates
in the Eucludean space, and $K=n+1$ for paraboloidal coordinates
in the Euclidean space.
The separation of variables (c.f.\cite{ku92,sk89}) is understood in
the context of the
given hierarchy of Hamiltonian systems as the construction of $K$
pairs of canonical variables $\pi_i,\,\mu_i,\,i=1,\ldots,K$,
\begin{equation}\{\mu_i,\mu_k\}=\{\pi_i,\pi_k\}=0,
\quad\{\pi_i,\mu_k\}=\delta_
{ik},\label{canvar}\end{equation}
and $K$ functions $\Phi_j$ such that
\begin{equation}
\Phi_j\left(\mu_j,\pi_j,H_N^{(1)},\ldots,H_N^{(K)}\right)=0,\quad
j=1,2,\ldots,K,\label{Phi}\end{equation}
where $H_N^{(i)}$ are the integrals of motion in involution.
The equations (\ref{Phi}) are the
{\it separation equations\,}. The integrable systems considered
admit the Lax representation in the form of
$2\times 2$ matrices (\ref{Lax}) and we will introduce the
separation variables $\pi_i,\mu_i$ as
\begin{equation}B(\mu_i)=0,\quad \pi_i=A(\mu_i),\quad i=1,\ldots,K.
\label{sepvar}
\end{equation}
Below we write explicitly these formulae for our systems.
The set of zeros $\mu_j,j=1,\ldots K$ of the function $B(u)$
defines the spherical
($\varepsilon=0$),
general ellipsoidal ($\varepsilon=1$) and general paraboloidal
($\varepsilon=4(u-x_{n+1}+B)$)
coordinates given by the formulae \cite{ka86,mi86,ku92}
\begin{eqnarray}
x_m^2&=&c{\prod_{j=1}^{n-1} (\mu_j-e_m)\over \prod_{k\neq m} (e_m-e_k)},
\quad
m=1,\ldots,n,\nonumber\\ &&{\rm where}\quad c=\sum_{k=1}^n x_k^2,
\quad{\rm for}\quad \varepsilon=0;\label{varsph} \\
x_m^2&=&-4{\prod_{j=1}^{n} (\mu_j-e_m)\over \prod_{k\neq m} (e_m-e_k)},
\quad
m=1,\ldots,n,\nonumber\\ &&\quad{\rm for}\quad \varepsilon=1;
\label{varell} \\
x_{n+1}&=&-\sum_{i=1}^n e_i +B +\sum_{i=1}^{n+1} \mu_i,
\nonumber\\ x_m^2
&=&-4{\prod_{j=1}^{n+1} (\mu_j-e_m)\over \prod_{k\neq m} (e_m-e_k)},
\quad
m=1,\ldots,n,\quad \nonumber \\ &&\quad{\rm for}
\quad \varepsilon=4(u-x_{n+1}+B).
\label{varpar}\end{eqnarray}
\begin{theorem} The coordinates $\mu_i,\pi_i$ given by
(\ref{sepvar}) are
canonically conjugated.\end{theorem}
{\bf Proof} Let us list the commutation relations between $B(v)$
and $A(u)$,
\begin{eqnarray}\{B(u),B(v)\}&=&\{A(u),A(v)\}=0,\label{uuvv}\\
\{A(u),B(v)\}&=&{2\over v-u}(B(u)-B(v)).\label{vuvu}\end{eqnarray}
The equalities $\{\mu_i,\mu_j\}=0$ follow from (\ref{uuvv}).
To derive the equality $\{\mu_i,\pi_j\}=-\delta_{ij}$ we substitute
$u=\mu_j$ in (\ref{vuvu}), obtaining thus
\[\{\pi_j,B(v)\}=-{2\over v-\mu_j}B(v),\]
which together with the equation
\[ 0= \{\pi_j,B(\mu_i)\}=
 \{\pi_j,B(v)\}\mid_{v=\mu_i}+B'(\mu_i)\{\pi_j,\mu_i\}\]
gives
\[\{\pi_j,\mu_i\} =-{1\over B'(\mu_i)}\{\pi_j,B(v)\}\mid_{v=
\mu_i}=\delta_{ij}.\]
Equalities $\{\pi_i,\pi_j\}=0$ can be verified in the similar way:
\begin{eqnarray}\{\pi_i,\pi_j\}&=&\{A(\mu_i),A(\mu_j)\}\nonumber\\
&=&\{A(\mu_j),A(v)\}\mid_{v=\mu_i}+A'(\mu_i)\{\mu_i,A(\mu_j)\}
\nonumber\\
&=&A'(\mu_j)\{A(\mu_i),\mu_j)\}+A'(\mu_i)\{\mu_i,A(\mu_j)\}=0.
\nonumber\end{eqnarray}

The separation equations have the form
\begin{equation}\pi_i^2=d(\mu_i),\label{sepeq}\end{equation}
where the function $d(u)$ is the determinant of the $L$-operator
(\ref{int}).

\subsection{Quantization}
The separation of variables has a direct quantum counterpart
\cite{kuz92,sk92a}. To pass
to quantum mechanics we change the variables $\pi_i,\mu_i$ to
operators and the Poisson brackets (\ref{canvar}) to the commutators
\begin{equation}[\mu_j,\mu_k]=[\pi_j,\pi_k]=0,
\quad[\pi_j,\mu_k]=-i\delta_{jk}.\label{qcanvar}
\end{equation}
Suppose that the common spectrum of $\mu_i$ is simple and the
momenta $\pi_i$
are realized as the derivatives
$\pi_j=-i{\partial\over \partial \mu_j}$. The separation equations
(\ref{sepeq}) become the operator equations, where the noncommuting
operators are assumed to be
ordered precisely in the order as those listed in (\ref{Phi}), that
 is $\pi_i,\mu_i,H_N^{(1)}$,
$\ldots,H_N^{(K)}$. Let $\Psi(\mu_1,\ldots,\mu_{K})$ be a common
eigenfunction of the
quantum integrals of motion:
\begin{equation}H_N^{(i)}\Psi=\lambda_i\Psi, \quad i=1,\ldots,K.
\end{equation}
Then the operator separation equations lead to the set of
 differential equations
\begin{equation}
\Phi_j(-i{\partial \over \partial \mu_j},\mu_j,H_N^{(1)},
\ldots,H_N^{(K)})\Psi(\mu_1,\ldots,\mu_{K})=0,
\quad j=1,\ldots,K,\label{oper}
\end{equation}
which allows the separation of variables
\begin{equation}
\Psi(\mu_1,\ldots,\mu_{K})=\prod_{j=1}^{K}\psi_j(\mu_j).
\end{equation}
The original multidimensional spectral problem is therefore reduced
to the set
of one-dimensional multiparametric spectral problems which have the
following
form in the context of the problems under consideration
\begin{eqnarray}&&\left({d^2\over d u^2}+\varepsilon u^N
+\sum_{i=1}^{n}{\lambda_i\over u-e_i}\right)
\psi_j(u;\lambda_1,\ldots,\lambda_n)=0,
\nonumber \\ &&{\rm for}\quad \varepsilon=0,1, \label{seqell} \\
&&\left({d^2\over d u^2}+16u^{N-2}(u+B)^2
+8\lambda_{n+1}+\sum_{i=1}^{n}{\lambda_i\over u-e_i}\right)
\psi_j(u;\lambda_1,\ldots,\lambda_{n+1})=0,\nonumber \\
&&{\rm for}\quad \varepsilon=4(u-x_{n+1}+B), \label{seqpar}\end{eqnarray}
with the spectral parameters $\lambda_1,\ldots,\lambda_{n+1}$.
The problems (\ref{seqell}),(\ref{seqpar}) must be solved on the
different intervals---``permitted zones"---for the variable $u$.

\section{Conclusion}
We remark that all systems considered yield the algebra which has
general properties being independent on the type of the system.
Therefore
it would be interesting to consider its Lie-algebraic origin within
the general approach to the classical $r$-matrices \cite{se85}.

There exists an interesting link of the algebra studied here
with the restricted flow formalism for the stationary flows of the
coupled KdV (cKdV) equations \cite{ar92}. The Lax pairs which have
been derived
in the paper from the algebraic point of view were recently found in
 \cite{arw92} by
consideration of the bi-Hamiltonian structure of cKdV.

It seems to be interesting to study the same questions for the
generalized
hierarchy of differential operators of Gelfand-Dickey for which the
corresponding $L$-operators have to be the $n\times n$ matrices.

\section*{Acknowledgements}
The authors are grateful to E K Sklyanin and A P  Fordy for valuable
discussions. We also
would like to acknowledge the EC for funding under the Science
programme SCI-0229-C89-100079/JU1.  One of us (JCE) is grateful to the
NATO Special Programme Panel on Chaos, Order and Patterns for support
for a collaborative programme, and to the SERC for research funding
under the Nonlinear System Initiative. VBK acknowledges support
from the National
Dutch Science Organization (NWO) under the Project \#611-306-540.

\end{document}